# Decoding University Hierarchy and Prestige in China through Domestic Ph.D. Hiring Network

**Authors:** Chaolin Tian[1†], Xunyi Jiang[1†], Yurui Huang[1], Langtian Ma[1], Yifang Ma[1*]

**Affiliations:**
[1]Department of Statistics and Data Science, Southern University of Science and Technology, Shenzhen, Guangdong, China
* Correspondence to: mayf@sustech.edu.cn
[†]These authors contribute equally to this work.

## Abstract

The academic job market for fresh Ph.D. students to pursue postdoctoral and junior faculty positions plays a crucial role in shaping the future orientations, developments, and status of the global academic system. In this work, we analyze the domestic Ph.D. hiring network of 23,994 scientists across Chinese Ph.D.-granting universities over three decades. We employ the Minimum Violation Rankings algorithm to decode the rankings for universities based on the Ph.D. hiring network, which offers a deep understanding of the structure and dynamics within the network. Our results uncover a consistent, highly structured hierarchy within this hiring network, indicating the imbalances wherein a limited number of universities serve as the main sources of fresh Ph.D. across diverse disciplines. Furthermore, over time, it has become increasingly challenging for Chinese Ph.D. graduates to secure positions at universities more prestigious than their alma maters. Drawing on Bourdieu's framework of academic capital and social reproduction, as well as Becher and Trowler's insights into academic tribes and disciplinary cultures, our study quantitatively situates these empirical findings within broader global processes of elite reproduction and highlights distinctive local factors in China's academic enterprise.

## 1. Introduction

Doctoral education serves an increasingly central role in the global academic ecosystem, with the growing number of Ph.D. candidates intensifying competition for faculty positions worldwide (Ghaffarzadegan et al. 2015; Nerad 2010). Research-focused doctoral training directly impacts the future standard of scientific inquiry, and a well-designed ecosystem for doctoral training and faculty recruitment is key to sustainable scientific development (McLoughlin, Richards, and Ivy 2019; Rybarczyk et al. 2011). When doctoral students seek academic employment after graduation, they base their decisions on factors including alma mater, field of study, mentorship, publication record, and university prestige. This selection process forms a dynamic network of doctoral student job placement. In this network, the nodes are universities or research institutes, and the edges represent the transition flow from the doctoral student's alma mater to their subsequent employing institutions.

The granting of doctoral degrees in China began as early as 1983 (Ma 2007), and over the past few decades, the number of doctoral students trained in China has steadily increased (Wenqin et al. 2018), significantly advancing the science and technology development in China. Despite the substantial growth in Ph.D. programs over the past five decades, Ph.D. graduates—who have invested significantly in their training to become researchers—face a saturated academic job market (Campbell and Hu 2010; Chen 2003). Two potential reasons account for this: firstly, the increasing number of Ph.D. candidates worldwide; and secondly, the influx of overseas doctoral students into Chinese universities, which has notably influenced the local doctoral employment market in universities (Wang, Tang, and Li 2015; Cao et al. 2020; Shi, Liu, and Wang 2023). The Ph.D. faculty hiring market is pivotal in shaping research priorities and the career trajectories of scholars (Clauset, Arbesman, and Larremore 2015).

Institutional prestige is a key determinant in this market. Prestige reflects the respect accorded by stakeholders and encompasses multiple characteristics of an institution (Topaler and Ayvaz-Çavdaroğlu 2022; Bloch and Mitterle 2017). Prior studies confirmed that the institutional prestige associated with scholars' doctoral degrees has consistently emerged as a primary determinant in shaping employment opportunities for individuals entering the academic labor market (Baldi 1995). Moreover, this prestige creates stratified hierarchies within the academic system—here, "hierarchy" refers to the stratification of universities resulting from long-standing prestige imbalances, whereby a small number of elite institutions command disproportionate resources, followed by mid-tier and lower-tier universities, respectively(Clauset, Arbesman, and Larremore 2015; Hanneman 2001; Burris 2004; Oprisko 2012). Wapman et al. expanded on this by revealing universal inequalities, wherein a small minority of universities disproportionately supplies the majority of faculty across various fields, and these disparities are accentuated by attrition patterns, underscoring steep hierarchies of prestige (Wapman et al. 2022). Furthermore, top prestigious universities often practice "social closure," dominating the majority flows in the Ph.D. placement network, a phenomenon facilitated by the relative long-term stability of these hierarchies (Fowler, Grofman, and Masuoka 2007; Kawakatsu et al. 2021).

However, there have been limited explorations in the existing body of literature to uncover the unique characteristics of China's domestic Ph.D. hiring network, which differs notably from other countries like the United States due to its unique social and economic structure, geographical distributions, policy system of education management, and the special historical and cultural traditions in the global environment (Lu and Zhang 2023; Jin and Horta 2018; Hertzum 2014). In this study, we empirically analyze the academic hiring network among universities in China, focusing on the transition from Ph.D. graduates to the next employed faculty positions. We address two main questions: (1) What are the structure and dynamics of the Ph.D. exchange network in China? and (2) How does institutional prestige relate to observed hiring patterns within this network?

Using dataset covering placement trajectories of 23,994 Ph.D. graduates from 1990 to 2020 across 501 Chinese universities or school-level academic units, we apply the network-based approach of Clauset et al. (Clauset, Arbesman, and Larremore 2015) to extract the prestige hierarchy structure among Chinese universities that best explain the observed Ph.D. hiring decisions. Our findings reveal a pronounced hierarchical structure with increasing difficulty over time for graduates to obtain positions at institutions more prestigious than their alma maters. These insights inform policy implications for Ph.D. placement, inequality, and career development of junior scientists.

We integrate two seminal theoretical frameworks to contextualize our results. Bourdieu's *The State Nobility* explains how elite institutions accumulate and convert academic capital to maintain dominance (Bourdieu 1998), while Becher and Trowler's *Academic Tribes and Territories* elucidates how disciplinary cultures shape career trajectories (Becher and Trowler 2001). This dual approach enriches our understanding of how China's academic hierarchies are reproduced both universally and locally.

## 2. Literature Review

### 2.1 Academic Careers, Faculty Mobility, and Power Structures

Academic careers are embedded within structured institutional and symbolic environments. As Bourdieu argues in The State Nobility, academic institutions operate as hierarchical fields in which symbolic capital, such as institutional prestige plays a central role in structuring opportunities for career advancement (Bourdieu 1998). Within this stratified space, the trajectory of Ph.D. graduates is shaped not only by individual attributes and scholarly outputs, but also by institutionalized mechanisms that reproduce power relations (Gerhards, Hans, and Drewski 2018). Becher and Trowler further emphasize the disciplinary differences in career norms and expectations, noting that each academic "tribe" adheres to distinct rules of mobility, prestige, and knowledge valuation (Becher and Trowler 2001).

A wide range of factors influence Ph.D. career trajectories, including personality, gender, mentorship, publication records, labor market competition, social capital, and financial considerations (Mangematin 2000; Kim, Benson, and Alhaddab 2018; Stephan 2012; Li et al. 2019; Li et al. 2020). While many Ph.D. holders initially aspire to secure academic positions, the persistent oversupply of candidates relative to available faculty roles—especially in research-intensive universities—has pushed an increasing number toward non-academic careers (Musselin 2007; Kassicieh and Nowak 1986; Cyranoski et al. 2011; Sauermann and Roach 2012; Gao et al. 2021). At the same time, the boundary between academic and non-academic employment is becoming more porous, as seen in the rise of hybrid roles in industry and applied research sectors (Stephan 2012; Roach and Sauermann 2010; Dietz and Bozeman 2005; Herrera and Nieto 2015; Jiang et al. 2024). These transitions, while often framed as individual choices, are conditioned by broader structural constraints — including prestige-based access to elite institutions and gatekeeping within academic disciplines.

Faculty mobility, both international and domestic, constitutes another mechanism through which academic capital circulates and is reallocated. Existing research has examined the determinants of such mobility, highlighting differences between transnational and intra-national flows. International mobility is often associated with national development levels, geopolitical dynamics, and immigration policy regimes(Marginson and van der Wende 2009; Mihut, de Gayardon, and Rudt 2016), whereas domestic mobility is more closely tied to research funding, institutional reputation, collaboration networks, and the availability of favorable working conditions (De Filippo, Casado, and Gómez 2009; Edler, Fier, and Grimpe 2011; Scellato, Franzoni, and Stephan 2017; Herrera, Muñoz-Doyague, and Nieto 2010; Lin, Frey, and Wu 2023). Empirical studies also show that mobility is associated with increased productivity, visibility, and collaboration, reinforcing the idea that mobility is both a cause and a consequence of academic success(Heitor, Horta, and Mendonça 2014; Wang et al. 2019).

In the Chinese context, faculty mobility patterns are shaped by both global academic norms and locally specific institutional dynamics. While structural determinants such as research funding and job prestige remain central, additional factors—such as age, hukou-based location preferences, and the infrastructural resources of host cities—play an important role (Liu, Cheung, and Hung 2022). For instance, early-career scholars under the age of 45 are more likely to relocate, yet faculty based in major academic hubs like

Beijing often choose to stay despite institutional dissatisfaction, due to the concentration of academic capital and professional networks(Yan, Yue, and Niu 2015; Liu et al. 2019). These findings reflect the embeddedness of faculty mobility within both institutional prestige hierarchies and regional opportunity structures.

Together, these strands of literature reveal that faculty careers are shaped by a complex interplay of individual, institutional, and disciplinary factors. Yet more importantly, they point to the operation of academic power structures that govern the distribution of opportunities in ways that reinforce existing hierarchies. By situating our empirical analysis within this broader theoretical context, we aim to illuminate how hiring patterns in Chinese academia reflect not just market dynamics, but also the ongoing reproduction of academic capital across institutional and disciplinary lines.

### 2.2 Theoretical Perspectives on Academic Hierarchy

Two seminal theoretical frameworks offer valuable insights into the mechanisms that drive academic hierarchies. By drawing on these frameworks, we can better understand how both universal processes and discipline-specific dynamics contribute to the observed patterns in faculty hiring.

**Bourdieu's Perspective:** Bourdieu's The State Nobility provides a compelling explanation of how elite institutions establish and maintain dominance (Bourdieu 1998). According to Bourdieu, these institutions accumulate and convert various forms of capital—symbolic, cultural, and economic—to secure their privileged positions within the academic field. His concept of social closure further explains how elite institutions deliberately restrict upward mobility by favoring individuals who already possess the necessary forms of capital. This theoretical stance is directly relevant to our following findings that very few Ph.D. graduates are able to secure positions at universities more prestigious than their own doctoral institutions, suggesting that the accumulation of academic capital plays a critical role in perpetuating hierarchical structures.

**Becher and Trowler's Perspective:** Complementing Bourdieu's analysis, Becher and Trowler's Academic Tribes and Territories emphasizes that academic disciplines operate as distinct communities—each with unique norms, values, and hiring practices (Becher and Trowler 2001). Particularly, the discussions in Chapters 5, 7, and 8 elucidate how these "tribes" develop their own internal hierarchies, which significantly influence faculty recruitment and career trajectories. The disciplinary variations in hiring patterns that we observe in our study—for instance, differences among material science, computer science, chemistry, and biology—can be effectively explained by the presence of these distinct academic cultures. This perspective highlights that beyond institutional prestige, the specific cultural practices and traditions within each academic discipline also shape the overall hierarchical organization of the field.

Together, Bourdieu's framework of academic capital accumulation and social closure, and Becher and Trowler's insights into disciplinary communities, provide a comprehensive theoretical basis for understanding the dynamics at play within the Chinese Ph.D. hiring network. These perspectives suggest that the persistent challenge for upward mobility among Ph.D. graduates is not merely a function of supply and demand but also a manifestation of deep-rooted social and cultural mechanisms that reinforce elite dominance across institutions and disciplines.

### 2.3 Prestige Ranking Methods

The examination of university prestige rankings stands as a crucial research domain, wielding direct influence over resource allocation, strategic planning, and decision-making within universities (Markovsky 2000). Analyzing faculty hiring networks grounded in institutional social network rankings serves not only to elucidate internal institutional relationships but also to unveil qualitative dimensions of social hierarchy (Hanneman 2001; Gayo-Avello 2013).

Prestige ranking methods encompass two dimensions, one of which involves assessing network analysis from the perspective of faculty placement. Various indicators grounded in social networks have been employed to construct prestige rankings. Out-degree is the simplest indicator of academic prestige, computed by tallying the number of faculty placements within each university (Katz et al. 2011; Hanneman 2001; Barnett and Feeley 2011; Yang et al. 2022). Alternative measurements include eigenvector centrality (Bonacich 2007), PageRank (Terviö 2011), closeness centrality (Barnett et al. 2010; Ran, Xu, and Jia 2024), etc. While rankings derived from these indicators may exhibit variation, they are generally related to those provided by entities such as the National Research Council or US News & World Report (Hanneman 2001; He et al. 2024).

The second dimension involves simultaneous consideration of both faculty placement and hiring to extract market patterns from the network (Katz et al. 2011; Myers, Mucha, and Porter 2011; Li et al. 2022; Ran et al. 2022). A university is deemed central if it places a significant number of graduates in other highly prestigious universities, whereas a prestigious university is one that hires faculty from highly reputed universities. The findings underscore the utility of prestige as a measure, revealing a positive correlation between higher prestige and increased reputation (Hanneman 2001; Clauset, Arbesman, and Larremore 2015).

## 3. Data and Methods

### 3.1 Ph.D. Hiring Network

Our primary datasets are sourced from ORCID (as of 2022) and OpenAlex (as of 2023). ORCID is an academic nonprofit organization that provides each researcher with a unique, persistent identifier (PID), enabling them to distinguish themselves from other researchers and track their activities, including education, employment, research work, and other academic endeavors. OpenAlex is a publicly accessible database that contains a vast collection of scholarly articles and detailed information for each paper, including authors, affiliations, publication dates, research fields, and citations (Priem, Piwowar, and Orr 2022). OpenAlex assigns each scholar a unique author ID along with her/his corresponding PID, which allows us to integrate the two databases and supplement missing author information in ORCID with the institutional details from OpenAlex. OpenAlex provides unique institution IDs and categorizes institutions into eight types: education, company, healthcare, government, facility, archive, nonprofit, and other.

In our study, we collected Ph.D. hiring records from ORCID, focusing on scholars who earned their doctorates in China and were subsequently employed by Chinese institutions. In detail, we collected the employment records, education records, and publication data of all scholars from ORCID—which provides each institution's RINGGOLD ID and each work's DOI—to link these records to OpenAlex, and then extracted each scholar's publication and citation records from OpenAlex, allowing us to accurately confirm their academic activities and performance. We further filter the academic institutions via OpenAlex's institution type "Education", removing other types like "company." Since ORCID is primarily utilized to track the academic research trajectories of authors based on their publications, our data includes more comprehensive recruitment information for STEM faculty.

Our integrated data encompass about 23,994 individuals who earned their Ph.D. degrees in Chinese universities and continued to conduct academic activities at Chinese institutions between 1990 and 2020, covering 501 academic units at the department or school level, with detailed information on each scholar's doctoral university, year of graduation, field of study (categorized according to the 19 discipline classifications in OpenAlex), and subsequent academic employment. In table S1 and table S2, we show the detailed statistics of our data in each discipline.

Finally, we constructed a directed Ph.D. faculty hiring network, represented by the $N \times N$ $(N = 501)$ adjacency matrix, $M$, where the nodes represent universities, and the weight of edge, $m_{ij}$ in $M$ is the number of Ph.D. graduated from university $i$ and being hired by university $j$.

### 3.2 Minimum Violation Rankings

If faculty placement conforms to a well-defined social hierarchy, individuals are unlikely to secure appointments at universities exceeding the prestige of their doctoral degree (Henrich and Gil-White 2001). The extent to which a particular faculty network aligns with this pattern can be gauged by identifying the minimum level of violation, signifying the closest proximity to this stringent criterion (Clauset, Arbesman, and Larremore 2015; De Vries 1998; Park 2010).

Within the framework of the faculty hiring network $M$, we utilize Minimum Violation Rankings (MVR) to measure the prestige of each university, capturing the hierarchy within the faculty hiring network (Clauset, Arbesman, and Larremore 2015). A violation arises when an edge $(i, j)$ is present, and the rank of $j$ surpasses that of $i$. MVR is specifically defined as a ranking on the vertices that minimizes the number of violations. Networks marked by robust prestige hierarchies will manifest only a limited count of such violations.

In faculty hiring network, each vertex signifies a university, and each directed edge $(i, j)$ denotes a faculty member at university $j$ who earned doctorate from university $i$. Let $\pi$ symbolize a permutation of the $n$ vertices within our directed network $M$, where edges carry positive weights and $\pi_i$ represents the rank of vertex $i$ in permutation $\pi$. A prestige hierarchy is then a ranking of vertices, where $\pi_i = 1$ signifies the highest-ranked vertex. The hierarchy's strength, denoted by $\rho$, is determined by the fraction of edges pointing downward, (i.e., those for which $\pi_i \leq \pi_j$), optimized across all rankings. Alternatively, $\rho$ represents the rate at which faculty are placed no higher in the hierarchy than their doctoral university. A value of $\rho = 1/2$ implies faculty move upward or downward in the hierarchy at equal rates, irrespective of their origin, while $\rho = 1$ indicates a flawless social hierarchy.

Let $\pi(M)$ represents an adjacent matrix for $M$, wherein rows and columns are reordered based on the permutation $\pi$. MVR is a permutation $\pi$ that minimizes the number of edges pointing up in the ranking or, equivalently, maximizes the number of edges pointing down. Determining such a permutation involves maximizing the net weight of unviolated edges, expressed as:

$$S_{\pi(M)} = \sum_{ij} M_{ij} \times sign[\pi(j) - \pi(i)],$$

which subtracts the weight of all rank-violating edges (found in the lower triangle, excluding the diagonal) of the reordered adjacency matrix $\pi(M)$ from the weight of non-violating edges. In the MVR model, to ensure validity, the model was strengthened by sampling from the posterior distribution and equilibrated the model with MCMC over 100,000 iterations to guarantee convergence. Subsequently, we explored rankings to minimize the occurrence of such violations, as outlined in the search procedures depicted in Figure 1. A low prestige score means a high prestige rank in the hierarchy of the faculty market.

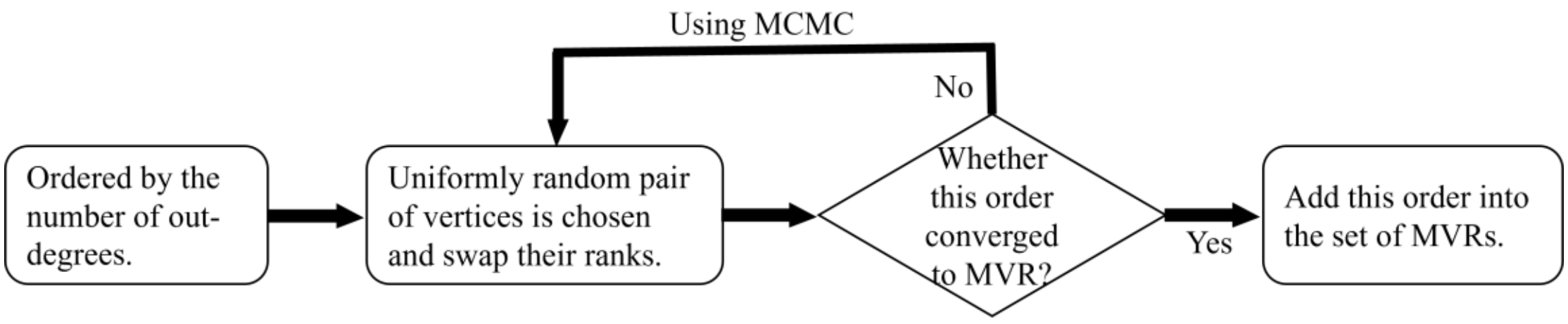

**Figure 1. The process of traversing to find the MVR** (Clauset, Arbesman, and Larremore 2015)**.** Given that multiple distinct vertex orderings can yield the same minimal number of violations, we employ a Markov chain Monte Carlo (MCMC) algorithm for sampling MVRs. The Metropolis-Hastings acceptance function with zero temperature is utilized for optimization. The process involves several steps. (1) Initial state: The ordering is based on the number of out-degrees. (2) Random selection: A pair of vertices are randomly chosen, their ranks are exchanged, and the violation edges are counted. (3) Acceptance criteria: If the violation of the new rank is less or equal to the previous rank, and the number of loops exceeds the burn-in steps, the MVR is added to the sample set. Otherwise, MCMC is used to repeat this step. (4) Mean rank computation: From the sampled rankings, the mean rank for each vertex is computed. (5) Prestige hierarchy creation: The sampled rankings with the highest $\rho$ are combined into a single prestige hierarchy. Each university is assigned a score (prestige score) equal to its average rank within the sampled set, and the order of these scores establishes the consensus ranking, a low prestige score means a high prestige rank in the hierarchy of the faculty network.

## 4. Results

### 4.1 Inequality in Chinese Ph.D. Hiring Network

Figure 2A displays the exponential growth trend of Chinese faculty production over the past three decades. From 1990 to 2017, the number of faculty members grew from 59 to 1,914 before slowing to 966 in 2020 (potentially due to incomplete data updates or the growing trend of hiring overseas graduates).

The Gini coefficient (G) is a widely recognized metric for measuring social inequality, defined as the average relative difference between observed values and a randomly selected pair of observations. G = 0 indicates perfect equality, whereas G = 1 indicates maximal inequality. Our analysis of faculty production (see Figure 2B) reveals a Gini coefficient of 0.78, indicating that 22% of Chinese universities account for nearly 78% of faculty output. This signifies a substantial concentration of Ph.D. faculty production within a small subset of universities in China, comparable to values of 0.62–0.76 reported for North America by Clauset et al. (Clauset, Arbesman, and Larremore 2015). Furthermore, as shown in Figure S2, the Gini coefficients have increased steadily over the past three decades: 0.63 (1990–2000), 0.73 (2001–2010), and 0.76 (2011–2020). These observations highlight that, alongside China's rapid scientific and technological development over the past three decades, the domestic faculty hiring hierarchy has become increasingly pronounced. The results reveal significant imbalances in the Chinese faculty production network, prompting us to quantify the extent of inequality in the entire faculty hiring network.

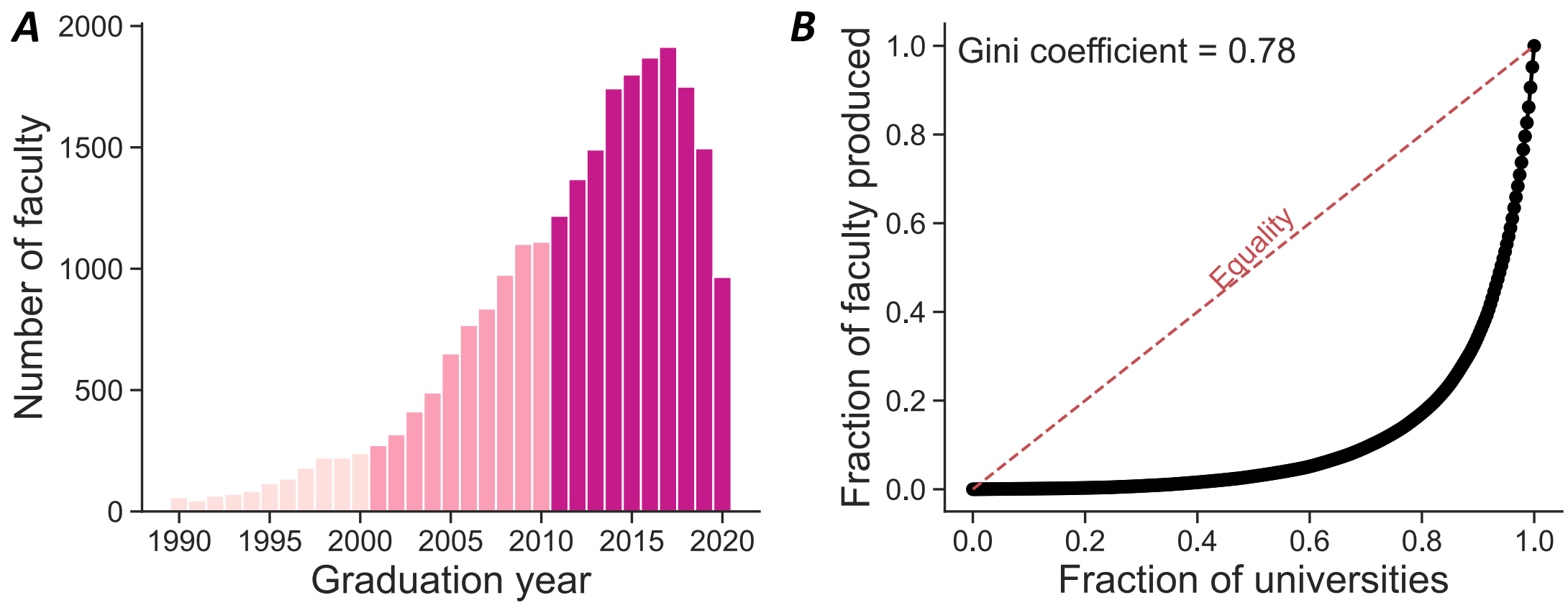

**Figure 2. Inequality in faculty production.** (A) The number of Chinese faculty has increased exponentially in three decades. (B) The Lorenz curve shows the fraction of all faculty produced as a function of producing universities, indicating a strong inequality in the output of faculty members.

### 4.2 University Hierarchical Structure and Prestige Rankings

To substantiate the observed inequality, we employed a null model that corroborates the hierarchical nature of the faculty hiring market (see Figure 3A). In detail, the MVR process shows that random graphs with skewed degree distributions can exhibit hierarchies with relatively few rank violations. We then assessed the statistical significance of the extracted MVRs by comparing the number of unviolated edges in the empirical data to that in a random graph with the same degree sequence. In this random-graph model, the skewness in the degree distribution is preserved, but the empirical structure of the prestige hierarchy is removed. Indeed, in Figure 3A, we found that the fraction of unviolated edges $\rho$ in the empirical data is significantly larger than the cases in the degree-preserving randomizations ($p < 10^{-5}$, *t*-test), and the figure clearly depicts the separation between the hierarchical signal produced by the null model and the empirical data.

To further explore and visualize the hierarchical structure, we applied nested Stochastic Block Models (SBMs). The SBMs provide a comprehensive and realistic approach to deciphering the hierarchical organization of networks and identifying underlying connectivity patterns (Peixoto 2019; Elliott et al. 2020; Peixoto 2014). This method models similarities and differences between universities as a graph, with edges representing relationships between node characteristics. The graph's community structure, revealing hierarchically organized patterns of similarity, is uncovered through stochastic block modeling. This process was applied separately to top-tier and other universities. Using nested SBMs, we categorized the universities in our dataset into four main blocks with distinct internal structures, ultimately forming nine subgroups with heterogeneous connectivity patterns (see Figure 3B).

The "C9 League" is an alliance of top-tier universities in China, comprising nine institutions: Peking University (PKU), Tsinghua University (THU), Zhejiang University (ZJU), Nanjing University (NJU), University of Science and Technology of China (USTC), Fudan University (FDU), Shanghai Jiao Tong University (SJTU), Harbin Institute of Technology (HIT), and Xi'an Jiaotong University (XJTU). Figure 3B shows that these "C9 League" universities form a distinct cluster, further verifying the hierarchical structure of the Chinese faculty hiring market and indicating that elite institutions of comparable prestige are frequently placed at similar levels.

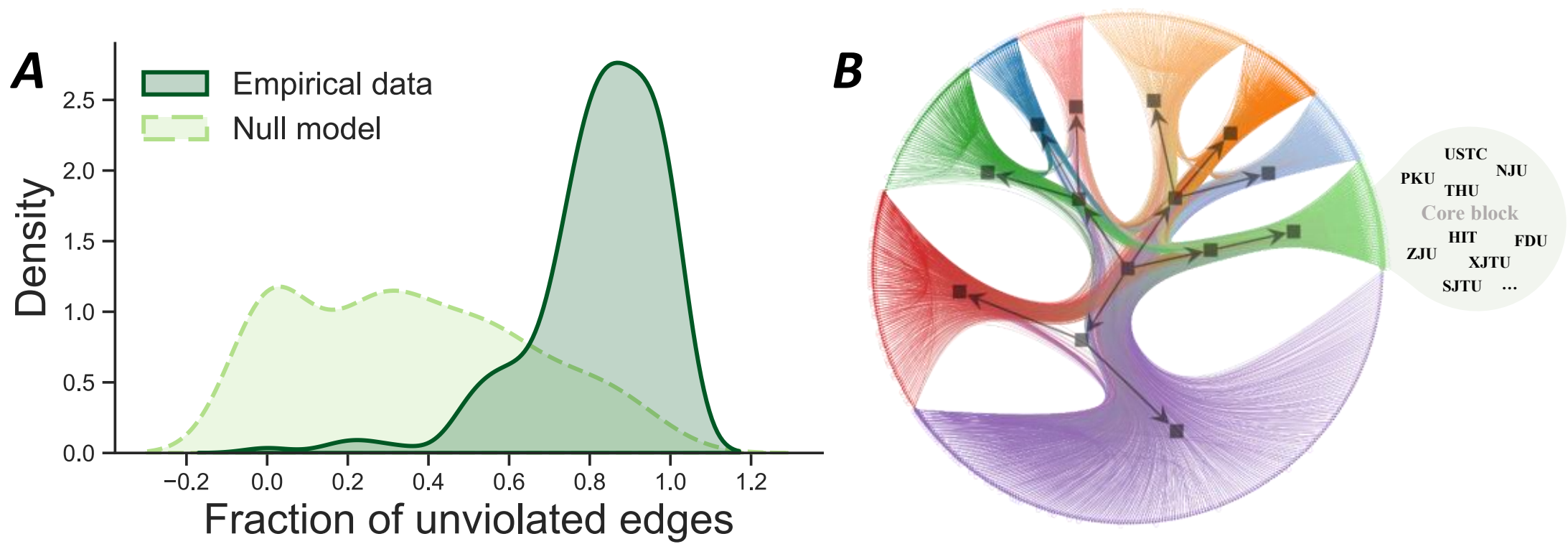

**Figure 3. Prestige hierarchies in faculty hiring network.** (A) Bootstrap distributions were generated to assess the fraction of unviolated edges $\rho$ both in the empirical data (depicted in dark green) and within a null model (illustrated in light green). The null model, designed to preserve the in- and out-degree sequences while randomizing connections, was employed for comparative analysis. The discernible gaps in the distributions signify the statistical significance of the extracted hierarchies. (B) The nested SBMs verify the hierarchical community structure (clustered black rectangles) of the faculty hiring market, displaying a prominent core-periphery architecture. The "blow up" shows the nodes that belong to the core block, containing "C9 League" universities. Each node signifies a university, the size of the nodes corresponds to the total degree, and the edge color indicates its direction (from dark to light).

In contrast to traditional university rankings, the ranking based on the faculty hiring network is consistent and further captures the latent prestige of universities in the domestic talent production and output market. Figure 4 presents the consensus rankings for the top 30 universities, highlighting the four disciplines with the highest faculty numbers: Material science, Computer science, Chemistry, and Biology.

Peking University and Tsinghua University consistently lead these rankings. Notably, we found that our MVR score captures information that other rankings do not, and therefore we compared our rankings with the QS Academic Reputation and Employer Reputation ranks (Figure S5).

First, Jilin University ranks higher in our MVR than in QS, reflecting its strong domestic reputation—being a top 985 institution located in Northeast China where Ph.D. graduates face less competition for academic positions. Second, we observe that specialized institutions such as Shanghai University of Traditional Chinese Medicine and Northwest A&F University, despite lower QS rankings, achieve higher MVR rankings due to their strong reputations in their respective fields. Finally, Wuhan University of Technology, although ranked highly by QS, is not part of the government's 985 Project; its domestic reputation is comparatively lower, resulting in a lower MVR ranking. This underscores the importance of 985 status in China's university evaluation.

In specific fields, Harbin Institute of Technology excels in computer science, China Agricultural University in chemistry, and Northwest A&F University in biology, each demonstrating notable standing. These findings suggest that, within the domestic Ph.D. faculty market, universities recognize and value each other's contributions to doctoral training in ways that conventional university rankings do not, indicating latent pipelines for talent cultivation and output. Furthermore, the difference between disciplines also highlights Becher and Trowler's perspective that academic disciplines function as distinct communities with unique norms, values, and power structures that influence hiring decisions (Becher and Trowler 2001).

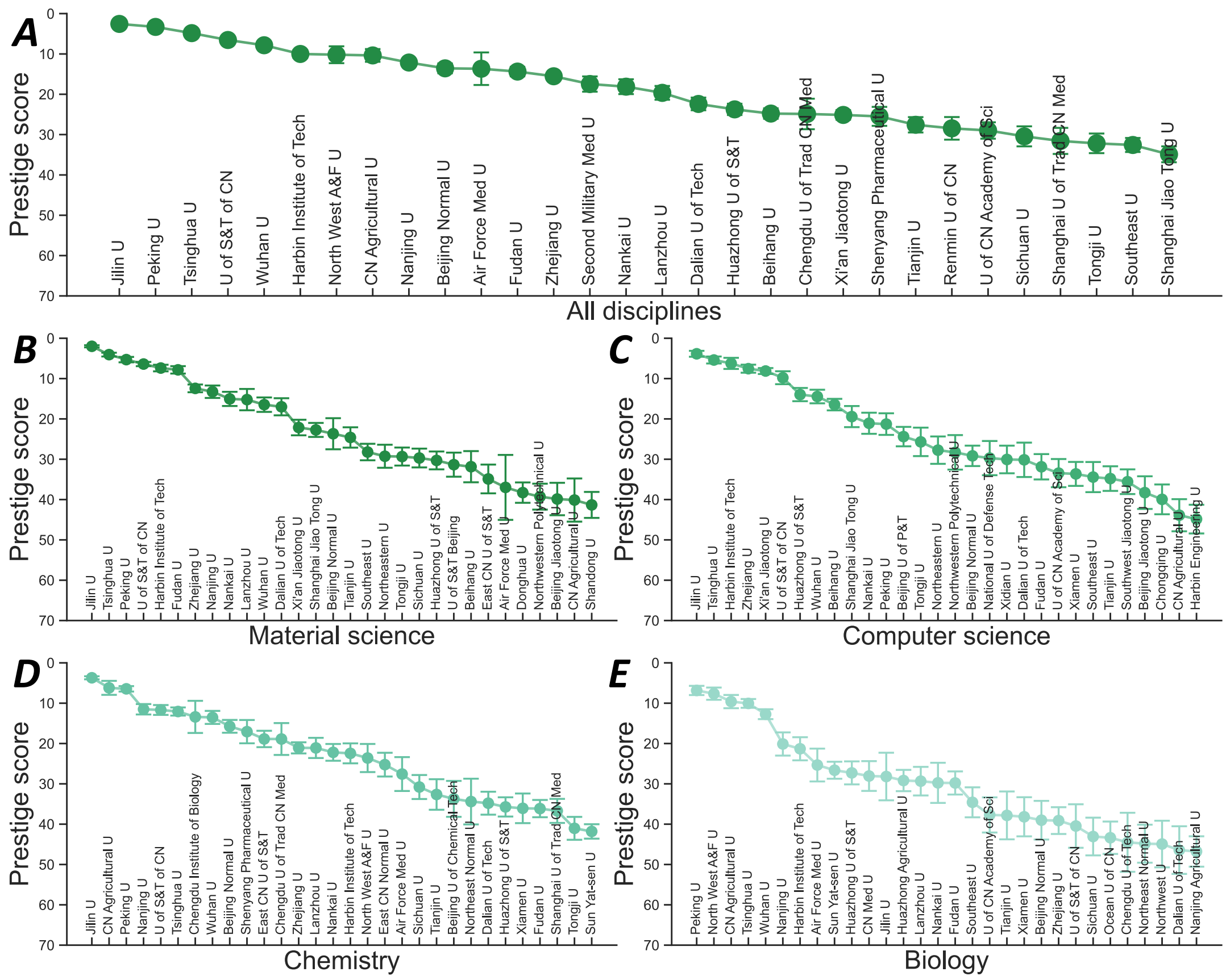

**Figure 4. Prestige rankings for the top 30 universities.** A low prestige score means a high prestige rank in the hierarchy of the faculty market. (A) All disciplines, (B) Material science, (C) Computer science, (D) Chemistry, (E) Biology. The nodes represent the means of the distribution for each university, and errors are at a 95% confidence interval. A line connects the means of each distribution. Abbreviations conventions: "CN" for China/Chinese, "U" for University, "S&T" for Science and Technology, "Sci" for Science, "Tech" for Technology, "A&F" for Agriculture and Forestry, "P&T" for Posts and Telecommunications, "Trad" for Traditional, "Med" for Medicine/Medical.

## 4.3 Escalating Challenges in Chinese Ph.D. Hiring Network

We delve into the temporal evolution of the hierarchy in the domestic Ph.D. hiring network. We divided the Ph.D. faculty into three cohorts according to their doctoral graduation years, each spanning a decade: 1990-2000, 2001-2010, and 2011-2020. Figure 5 illustrates the prestige of the top 10 universities during the three periods. We found that, over time, the hierarchy in the faculty hiring market has become increasingly rigid. For example, during the 1990-2000 period, the top three universities, as measured by prestige, were Jilin University, Nanjing University, and Xi'an Jiaotong University, with their prestige scores around 10 to 30, indicating a less pronounced hierarchy in the early stages of Chinese faculty hiring market (see Figure 5A). In other words, during this period, doctoral students from lower-prestige universities still had viable opportunities to secure positions at higher-prestige institutions. However, as time progressed, the prestige scores of leading universities have been approaching 0 significantly. This trend is particularly evident in the 2011-2020 period, where Tsinghua University and Peking University had nearly identical prestige scores that far exceeded those of other institutions (see Figure 5C). The increasing prestige of these top-tier universities correlates with their increased Ph.D. production and better placement opportunities, which in turn enhances their dominant position in the market and further intensifies the hierarchy.

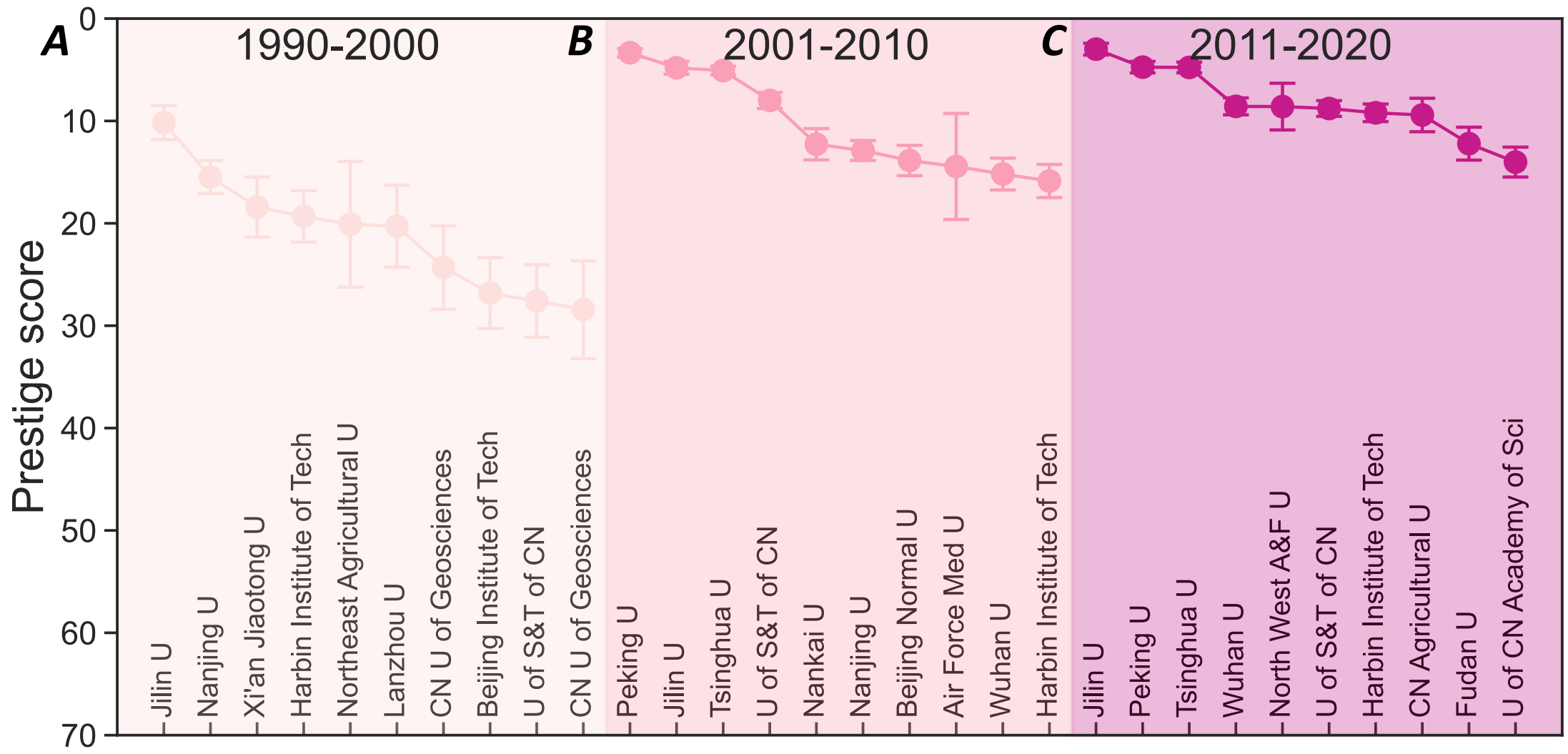

**Figure 5. Prestige scores for the top 10 universities of different periods.** (A) 1990-2000; (B) 2001-2010; (C) 2011-2020. A low prestige score means a high prestige rank in the hierarchy of the faculty market. The nodes represent the means of the distribution for each university, and errors are at a 95% confidence interval. A line connects the means of each distribution. Abbreviations conventions: "CN" for China/Chinese, "U" for University, "S&T" for Science and Technology, "Sci" for Science, "Tech" for Technology, and "Med" for Medicine/Medical.

Figure 6 presents the relative change in rank from doctoral universities to the universities where each Ph.D. graduate is subsequently placed, across four disciplines and over time. Our findings indicate a pronounced prestige hierarchy, with only 9.3% of faculty securing positions at universities more prestigious than their doctoral alma maters (see Figure 6A). By discipline, the numbers are 9.1% in material science, 9.3% in computer science, 8.8% in chemistry, and 10.1% in biology (see Figure 6A insert map). This result suggests that doctoral students in the field of biology are more likely to find prestigious faculty-hired universities compared to their doctoral schools, whereas it is most challenging for those in chemistry.

Over different decades, 9.9% of domestic Ph.D.s in the 1990-2000 cohort found positions more prestigious than their doctoral alma mater. In the 2001-2010 cohort, this percentage dropped to 8%. The relative decline from 9.9% to 8% represents a 23.8% decrease [(9.9 to 8%)/8%], indicating a lower prestige of placement universities in the 2001-2010 period compared to the 1990–2000 period. Further, the average relative rank-change score in 2001-2010 is 0.19 (19%), higher than the 0.17 (17%) observed in 1990-2000, revealing a 12.7% increase in rank change [(19% -17%)/17%].This shift is visually represented in Figure 6B, where the right-tail of the distribution for the 2001-2010 period extends further to the right, suggesting an increasing trend of Ph.D. graduates securing lower-prestige positions. A similar pattern is observed when comparing 1990-2000 and 2011-2020. This aligns with Bourdieu's perspective that elite institutions establish and maintain dominance by accumulating and converting various forms of capital, as well as how social closure restricts upward mobility, suggesting that the accumulation of academic capital plays a critical role in perpetuating hierarchical structures (Bourdieu 1998).

Over time, there is a gradual rightward shift in the distribution of prestige score change rank rates, denoting an expanding gap between the rankings of universities and the doctoral schools where Ph.D. graduates are hired. This pattern underscores a progressive tightening of the hierarchy within China's faculty market over time.

In summary, a hierarchical structure exists in the Chinese academic market, varying across different fields, with chemistry showing the most pronounced hierarchy. As time progresses, this hierarchical phenomenon intensifies, indicating that it is becoming increasingly challenging for Ph.D. graduates to advance upwards in the academic hierarchy, often leading them to accept positions at universities significantly lower in prestige than their doctoral universities.

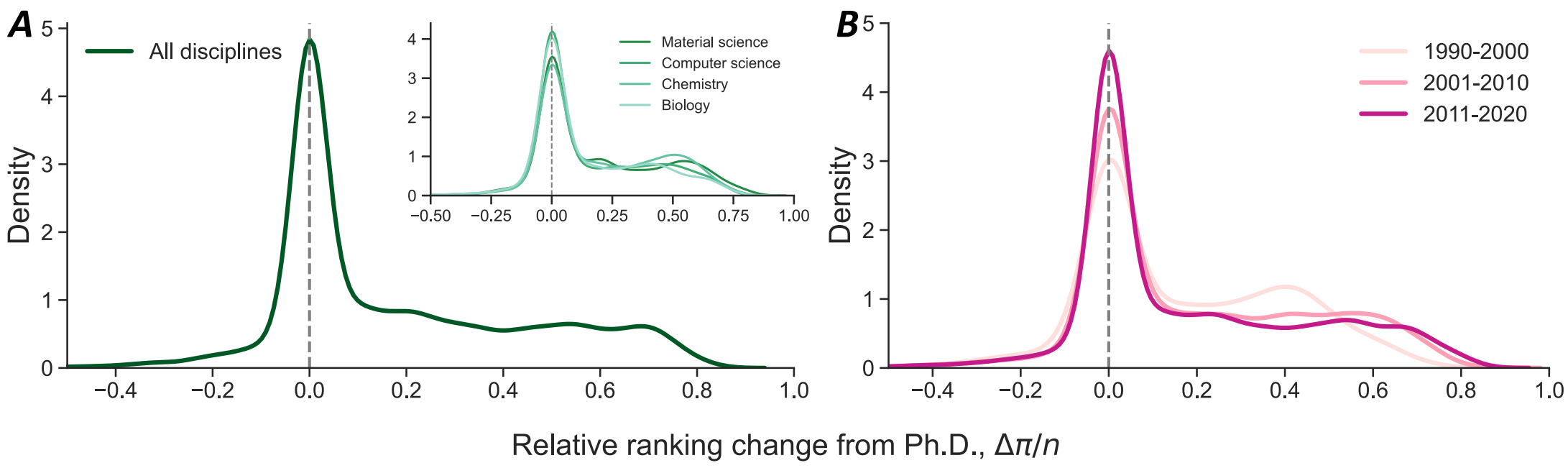

**Figure 6. Relative ranking changes from doctoral to current university.** (A) All disciplines, and the inset map shows the distribution of four disciplines: Material science, Computer science, Chemistry, and Biology. (B) Relative ranking changes across the years. The two-sided Kolmogorov–Smirnov test indicates that we can reject the null hypothesis in which the two distributions are drawn from different distributions ($p<0.001$ for pairwise of these three periods).

### 4.4 The Role of Postdoctoral Experience, Scientific Performance, and University Prestige

In recent decades, China's science and technology sector has grown rapidly, evidenced by an increasing number of publications and researchers. Meanwhile, China has introduced multiple overseas talent recruitment and incentive policies, including the Thousand Talents Plan. These initiatives have intensified competition for domestic faculty positions: over the past ten years, many universities have found it increasingly difficult to recruit Ph.D.-qualified faculty, with numerous candidates—many holding one or more postdoctoral appointments—competing for each position. Universities have accordingly raised their academic requirements, significantly impacting domestically trained Ph.D. graduates and reducing their chances of securing positions at universities ranked above their alma maters (i.e., upward placement).

In light of these dynamics, we examine the factors that influence whether domestic Ph.D. graduates secure faculty positions at universities more prestigious than their doctoral alma maters. To this end, we specify five nested models, each progressively incorporating key explanatory variables. The full model is given by:

$$Y_{it} = \beta_0 + \sum_k \beta_k X_{kit} + \gamma_{d(i)} + \delta_{g(i)} + \theta_{t(i)} + \varphi_{u(i)} + \phi_{r(i)} + \epsilon_{it},$$

where $Y_{it}$ is a binary indicator for upward placement for Ph.D. $i$ at relative year $t$ with $t$ denotes the number of years prior to securing a faculty position).

$X_{kit}$ comprises stepwise-added explanatory variables reflecting individual, institutional, and regional characteristics. Binary indicators include: (1) whether Ph.D. $i$ had postdoctoral experience before faculty placement; (2) whether their doctoral institution is affiliated with Project 985, a national initiative for elite universities; and (3) whether it is a newly established university founded within the past 30 years. Continuous controls capture research productivity (log-transformed number of publications and citations), collaboration scale (average number of coauthors per paper), and regional context (log-transformed GDP per capita of the province hosting the doctoral institution). The models include the following fixed effects: $\gamma_{d(i)}$ denotes for discipline; $\delta_{g(i)}$ denotes for graduation cohort (categorized into three graduation cohorts: Ph.D. graduating in 1990-2000, 2001-2010, and 2011-2020); $\theta_{t(i)}$ denotes relative placement year; $\varphi_{u(i)}$ denotes university type (including A&F, Medical, and others); $\phi_{r(i)}$ denotes Ph.D.'s doctoral alma mater geographic region (including East, Center, Northeast, and West of China). $\epsilon_{it}$ is the error term.

**Table 1. Regression models predicting upward placement in faculty hiring.** Models 1-5 sequentially introduce institutional characteristics (e.g., Project 985 and new university dummy), research productivity indicators (number of papers and citations, log-transformed), and control variables including team size and regional GDP. All models include fixed effects for university type (including A&F, Medical, and others), geographic region (including East, Center, Northeast, and West), discipline, graduation cohort, and relative placement year. Robust standard errors clustered at the Ph.D.'s level are reported in parentheses.

| | **Upward placement** | | | | |
|---|---|---|---|---|---|
| | **Model 1** | **Model 2** | **Model 3** | **Model 4** | **Model 5** |
| **Postdoctoral experience** | 0.025*** | -- | -- | -- | -- |
| | (0.008) | -- | -- | -- | -- |
| **Project 985** | -- | -0.133*** | -- | -- | -- |
| | -- | (0.013) | -- | -- | -- |
| **New university** | -- | -- | 0.144*** | -- | -- |
| | -- | -- | (0.047) | -- | -- |
| **Log(#Papers)** | -- | -- | -- | 0.074*** | -- |
| | -- | -- | -- | (0.011) | -- |
| **Log(#Citations)** | -- | -- | -- | -- | 0.017*** |
| | -- | -- | -- | -- | (0.004) |
| **Team size** | -0.001 | <0.001 | -0.001 | -0.002 | -0.001 |
| | (0.001) | (0.001) | (0.001) | (0.001) | (0.001) |
| **Log(GDP)** | -0.051* | -0.039* | -0.055** | -0.049* | -0.049* |
| | (0.019) | (0.019) | (0.019) | (0.019) | (0.019) |
| **University type** | Y | Y | Y | Y | Y |
| **Region** | Y | Y | Y | Y | Y |
| **Discipline** | Y | Y | Y | Y | Y |
| **Graduation cohort** | Y | Y | Y | Y | Y |
| **Relative placement year** | Y | Y | Y | Y | Y |
| **Constant** | 0.321*** | 0.380*** | 0.347*** | 0.287** | 0.305*** |
| | (0.081) | (0.081) | (0.081) | (0.082) | (0.082) |
| **Observations** | 56,419 | 56,419 | 56,419 | 56,419 | 56,419 |
| **R-squared** | 0.021 | 0.043 | 0.022 | 0.024 | 0.021 |

Robust standard errors in parentheses. *** p<0.001, ** p<0.01, * p<0.05

As shown in Table 1, postdoctoral experience has a significant positive effect on upward placement (coefficient = 0.025, $p < 0.001$; Model 1), suggesting that individuals with postdoctoral training are more likely to secure positions at more prestigious institutions than their doctoral alma maters. Research productivity also plays a key role: both the number of papers (coefficient = 0.074, $p < 0.001$; Model 4) and citation impact (coefficient = 0.017, $p < 0.001$; Model 5) are positively associated with upward mobility. Graduates from newly established doctoral institutions exhibit a higher probability of upward placement (coefficient = 0.144, $p < 0.001$; Model 3), whereas holding a Ph.D. from a Project 985 university is negatively associated with upward mobility (coefficient = -0.133, $p < 0.001$; Model 2), likely because such graduates already originate from the top of the prestige hierarchy. In addition, being located in a region with a higher GDP is associated with a lower likelihood of upward placement (coefficients ranging from -0.049 to -0.051, $p < 0.05$ across all models), possibly reflecting intensified competition and the growing preference of elite institutions in economically developed areas for overseas-trained Ph.D. holders.

These results highlight the dual influence of individual credentials, such as postdoctoral training and research output and structural constraints rooted in institutional prestige and regional context. While research merit facilitates upward movement, the reproduction of hierarchical hiring patterns remains strongly shaped by existing power structures in the academic field.

## 5. Discussion

Overall, our analysis covers the academic employment of nearly 24,000 fresh Ph.D. graduates who obtained faculty positions at universities, and the data span elite Ph.D.-granting Chinese institutions over three decades. We reveal distinct patterns and hierarchies in Chinese faculty hiring networks. The hiring network exhibits a heterogeneous out-degree distribution, indicating a strong inequality in faculty production. Our findings indicate that the Chinese faculty hiring market follows a steep hierarchical structure, dominated by a small minority of universities that train the majority of faculty members and occupy the highest levels of prestige, and this trend has become increasingly pronounced in recent decades. The inequality in faculty production in China is comparable to that observed in the North American context (Clauset, Arbesman, and Larremore 2015). Further, this hierarchical structure has gained increased prominence within the faculty market over time.

Currently, while the Chinese government has sought to alleviate employment pressures by expanding graduate enrollment, the appeal of academic positions for highly educated individuals has diminished, driving more doctoral graduates toward positions at lower-prestige universities. Our results confirm this phenomenon by revealing that only 9.3% of faculty secure placements at institutions more prestigious than their doctoral alma maters. This finding corroborates Bourdieu's theoretical argument that elite institutions leverage accumulated academic capital to maintain their dominant positions, effectively constraining upward mobility (Bourdieu 1998). Furthermore, the variations in hiring patterns across disciplines support Becher and Trowler's notion that distinct disciplinary cultures—each with their own norms and values—play a significant role in shaping academic career trajectories (Becher and Trowler 2001). These disciplinary "tribes" appear to modulate the overall hierarchy, as evidenced by the differences observed among fields such as chemistry, biology, computer science, and material science.

Our dual observations, which couple Bourdieu's insights on social reproduction with Becher and Trowler's understanding of academic tribes, demonstrate that China's academic hierarchies are shaped by both universal elite-reproduction processes and distinctive local and disciplinary factors that amplify market stratification. However, the rapid roll-out of "985/211" and "Double First Class" initiatives, the rise of postdoctoral training, and the regional diversification of new universities all challenge and extend Bourdieu's and Becher & Trowler's rules—newer institutions sometimes outperform older elites in upward mobility, revealing a dynamic interplay between institutional age and symbolic capital that is less evident in Western settings.

Our analysis suggests that many inequalities at the faculty hiring stage are later magnified over time, reflecting profound social inequalities. Moreover, the deepening of China's internationalization process, characterized by a growing number of Ph.D. graduates returning to work after studying abroad, is poised to influence the faculty market (Cao et al. 2020; Shi, Liu, and Wang 2023; Wang, Tang, and Li 2015), calling a more inclusive policy that balance talents from domestic and overseas universities. These findings suggest several strategic actions: Given the growing number of Ph.D. graduates seeking faculty positions, policymakers should expand postdoctoral opportunities to alleviate market saturation and ensure balanced prospects for foreign-trained and domestically trained scientists. At the same time, targeted grants and mentorship for newly founded institutions can help them build reputational capital and compete in the talent market. Monitoring disciplinary hiring disparities and directing resources to fields with the steepest hierarchies will foster more equitable growth, while regionally tailored incentives (e.g., relocation allowances) can mitigate geographic imbalances. Notably, China's rapid implementation of the "985/211" and "Double First Class" initiatives has accelerated symbolic capital accumulation for select universities, and the increasing prominence of postdoctoral training has reshaped credential norms. The fact that newer, regionally diverse universities sometimes outperform older elites in upward mobility underscores that prestige can be actively constructed through institutional entrepreneurship and supportive policy frameworks, rather than merely inherited.

Our research also has certain limitations that warrant future consideration. First, the MVR model used to evaluate prestige solely considers the number of mobile Ph.D.s between universities. However, the prestige of a university is also influenced by research conditions, salaries, facilities, and geographic location, which can be reflected in postdoctoral research outputs and faculty turnover rates (Yu, Szymanski, and Jia 2021). Indeed, in our regression model encompassing the university's reputation classification (Project 985), scholars' scientific performance and geographical characteristics, we found that these factors could affect the scholar's job placement to a highly ranked university than her/his alma maters. Future attempts would integrate these factors into the MVR model. Further, ORCID primarily tracks Ph.D. movements based on research publications and self-reported information, which may underrepresent scholars who are less active in publishing or who do not maintain comprehensive profiles. Addressing this bias in future studies—through enhanced data acquisition and integration—can improve the robustness of our conclusions.

By quantitatively analyzing the dynamics of faculty hiring in Chinese academia, our research provides valuable insights into the organization, composition, and scholarly contributions of the academic workforce. These findings can inform policies and strategies aimed at enhancing diversity, equity, and fairness in faculty recruitment and career development within Chinese universities. Ultimately, our study contributes to a deeper understanding of the faculty hiring market and its implications for the Chinese academic landscape.

## Figures legends

**Figure 1. The process of traversing to find the MVR.** Given that multiple distinct vertex orderings can yield the same minimal number of violations, we employ a Markov chain Monte Carlo (MCMC) algorithm for sampling MVRs. The Metropolis-Hastings acceptance function with zero temperature is utilized for optimization. The process involves several steps. (1) Initial state: The ordering is based on the number of out-degrees. (2) Random selection: A pair of vertices are randomly chosen, their ranks are exchanged, and the violation edges are counted. (3) Acceptance criteria: If the violation of the new rank is less or equal to the previous rank, and the number of loops exceeds the burn-in steps, the MVR is added to the sample set. Otherwise, MCMC is used to repeat this step. (4) Mean rank computation: From the sampled rankings, the mean rank for each vertex is computed. (5) Prestige hierarchy creation: The sampled rankings with the highest $\rho$ are combined into a single prestige hierarchy. Each university is assigned a score (prestige score) equal to its average rank within the sampled set, and the order of these scores establishes the consensus ranking, a low prestige score means a high prestige rank in the hierarchy of the faculty network.

**Figure 2. Inequality in faculty production.** (A) The number of Chinese faculty has increased exponentially in three decades. (B) The Lorenz curve shows the fraction of all faculty produced as a function of producing universities, indicating a strong inequality in the output of faculty members.

**Figure 3. Prestige hierarchies in faculty hiring network.** (A) Bootstrap distributions were generated to assess the fraction of unviolated edges $\rho$ both in the empirical data (depicted in dark green) and within a null model (illustrated in light green). The null model, designed to preserve the in- and out-degree sequences while randomizing connections, was employed for comparative analysis. The discernible gaps in the distributions signify the statistical significance of the extracted hierarchies. (B) The nested SBMs verify the hierarchical community structure (clustered black rectangles) of the faculty hiring market, displaying a prominent core-periphery architecture. The “blow up” shows the nodes that belong to the core block, containing “C9 League” universities. Each node signifies a university, the size of the nodes corresponds to the total degree, and the edge color indicates its direction (from dark to light).

**Figure 4. Prestige rankings for the top 30 universities.** A low prestige score means a high prestige rank in the hierarchy of the faculty market. (A) All disciplines, (B) Material science, (C) Computer science, (D) Chemistry, (E) Biology. The nodes represent the means of the distribution for each university, and errors are at a 95% confidence interval. A line connects the means of each distribution. Abbreviations conventions: “CN” for China/Chinese, “U” for University, “S&T” for Science and Technology, “Sci” for Science, “Tech” for Technology, “A&F” for Agriculture and Forestry, “P&T” for Posts and Telecommunications, “Trad” for Traditional, “Med” for Medicine/Medical.

**Figure 5. Prestige scores for the top 10 universities of different periods.** (A) 1990-2000; (B) 2001-2010; (C) 2011-2020. A low prestige score means a high prestige rank in the hierarchy of the faculty market. The nodes represent the means of the distribution for each university, and errors are at a 95% confidence interval. A line connects the means of each distribution. Abbreviations conventions: “CN” for China/Chinese, “U” for University, “S&T” for Science and Technology, “Sci” for Science, “Tech” for Technology, and “Med” for Medicine/Medical.

**Figure 6. Relative ranking changes from doctoral to current university.** (A) All disciplines, and the inset map shows the distribution of four disciplines: Material science, Computer science, Chemistry, and Biology. (B) Relative ranking changes across the years. The two-sided Kolmogorov–Smirnov test indicates that we can reject the null hypothesis in which the two distributions are drawn from different distributions ($p<0.001$ for pairwise of these three periods).

**Table 1. Regression models predicting upward placement in faculty hiring.** Models 1-5 sequentially introduce institutional characteristics (e.g., Project 985 and new university dummy), research productivity indicators (number of papers and citations, log-transformed), and control variables including team size and regional GDP. All models include fixed effects for university type (including A&F, Medical, and others), geographic region (including East, Center, Northeast, and West), discipline, graduation cohort, and relative placement year. Robust standard errors clustered at the Ph.D.'s level are reported in parentheses.

## Data availability

The database we used in this work (ORCID: https://orcid.org/; OpenAlex: https://openalex.org/) is publicly available. The datasets generated during the current study are available from the corresponding author on reasonable request.

## Competing interests

The authors have no competing interests.

## Ethical statements

This article does not contain any studies with human participants performed by any of the authors.

## Informed consent

This research does not involve human participants or animals.

# Support Information for text "Decoding University Hierarchy and Prestige in China through Domestic Ph.D. Hiring Network"

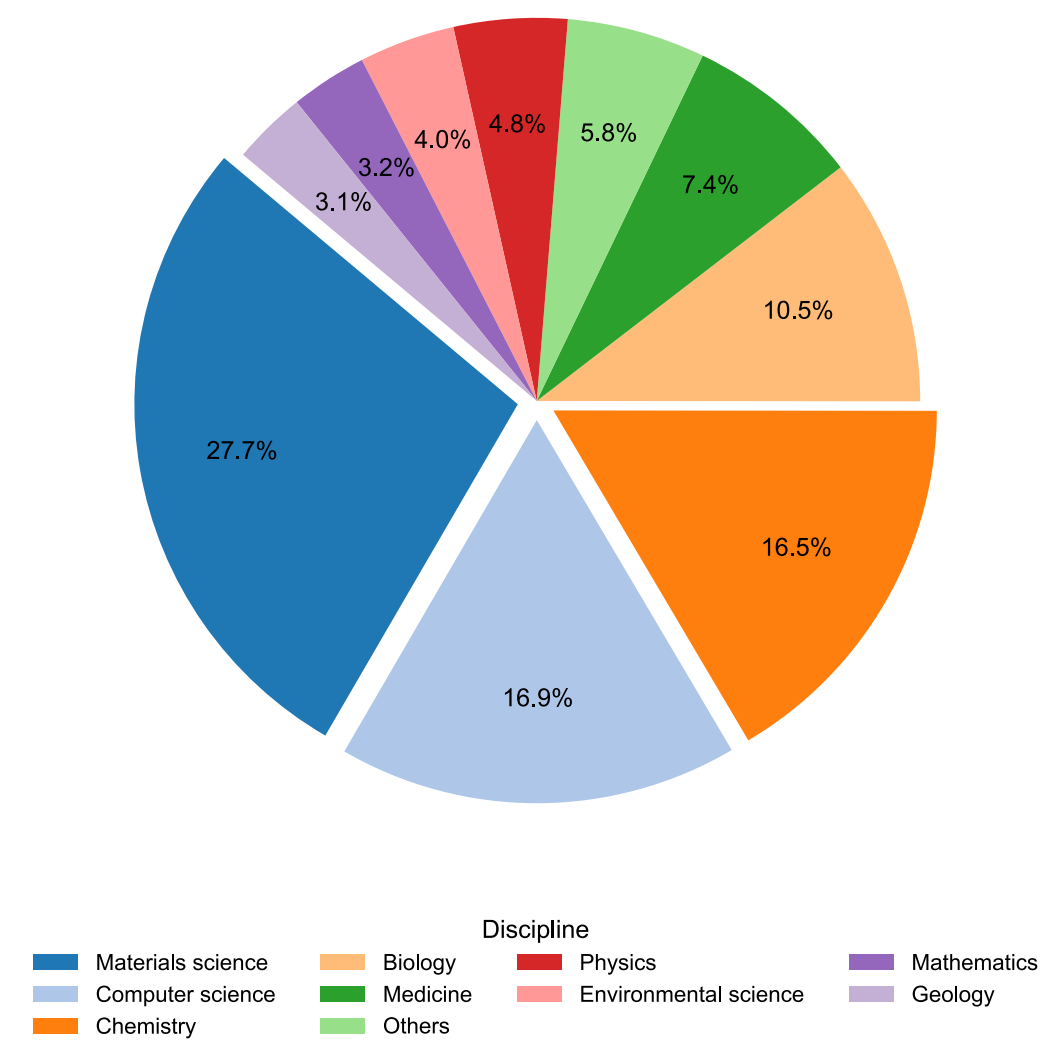

**Figure S1. Disciplinary distribution and prevalence of postdoctoral experience among faculty with domestic Ph.D. degrees in China.** Distribution of faculty by discipline among Ph.D. holders in our dataset. The majority of faculty come from STEM fields, with Materials Science (27.7%), Computer Science (16.9%), and Chemistry (16.6%) being the most represented.

**Table S1. Descriptive statistics of Ph.D. faculty by discipline.** This table summarizes the distribution of Ph.D. faculty across academic disciplines in the dataset, including the number and percentage of Ph.D.s, the proportion with postdoctoral experience, and the share of graduates from institutions affiliated with Project 985, Project 211, and China's national "first-class university" initiative. Materials science, computer science, and chemistry represent the largest fields by Ph.D. count, while disciplines such as philosophy, art, and history are sparsely represented. Substantial variation exists across fields in postdoctoral prevalence and institutional prestige backgrounds.

| Discipline | # of Ph.D.s | % of total Ph.D.s | Proportion with postdoc experience | Proportion from project 985 universities |
|---|---|---|---|---|
| Materials science | 6318 | 27.74% | 0.37 | 0.81 |
| Computer science | 3840 | 16.86% | 0.27 | 0.8 |
| Chemistry | 3757 | 16.50% | 0.41 | 0.77 |
| Biology | 2381 | 10.45% | 0.45 | 0.66 |
| Medicine | 1694 | 7.44% | 0.27 | 0.63 |
| Physics | 1090 | 4.79% | 0.49 | 0.88 |
| Environmental science | 921 | 4.04% | 0.37 | 0.77 |
| Mathematics | 736 | 3.23% | 0.26 | 0.8 |
| Geology | 711 | 3.12% | 0.35 | 0.58 |
| Others | 1328 | 5.83% | 0.26 | 0.75 |

**Table S2. Research productivity and team size of Ph.D. faculty before placement, by discipline.** This table reports the average number of publications, average citations per publication, and average team size (number of coauthors per paper) for each Ph.D. faculty member, separately measured for the five years before graduation and the fifteen years after graduation. Substantial variation is observed across disciplines in both pre- and post-graduation scholarly productivity. Values in parentheses indicate the standard errors of the means.

| Discipline | Before placement | | |
|---|---|---|---|
| | #Papers | #Cumulative citations | Average team size |
| Materials science | 5.32 (0.11) | 41.72 (0.96) | 6.11 (0.03) |
| Computer science | 5.21 (0.17) | 19.55 (0.76) | 4.48 (0.04) |
| Chemistry | 4.27 (0.11) | 41.44 (1.38) | 6.60 (0.06) |
| Biology | 3.39 (0.09) | 47.56 (2.46) | 8.16 (0.12) |
| Medicine | 4.56 (0.22) | 29.83 (1.53) | 8.09 (0.15) |
| Physics | 4.94 (0.19) | 21.89 (1.46) | 7.25 (0.39) |
| Environmental science | 4.36 (0.20) | 36.72 (2.15) | 6.47 (0.22) |
| Mathematics | 3.54 (0.25) | 14.53 (1.47) | 2.91 (0.07) |
| Geology | 3.54 (0.21) | 25.93 (1.70) | 5.35 (0.13) |
| Others | 3.53 (0.19) | 25.69 (2.63) | 5.07 (0.28) |

Values in parentheses indicate the standard errors of the means.

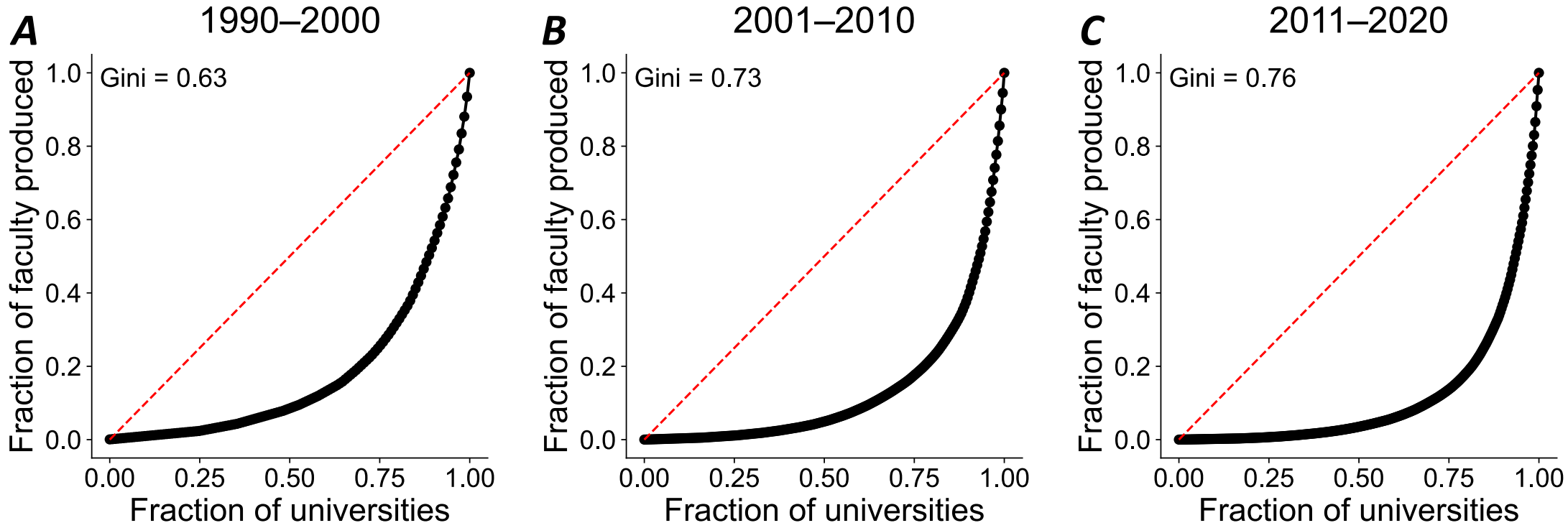

**Figure S2. Increasing concentration of faculty production across Chinese universities over time.** Lorenz curves show the cumulative distribution of faculty production across universities for three graduation cohorts:(A) 1990 – 2000, (B) 2001 – 2010, and (C) 2011 – 2020.The x-axis represents the cumulative fraction of institutions (ordered by faculty output), while the y-axis shows the cumulative fraction of Ph.D. faculty produced. The red dashed line denotes perfect equality. The Gini coefficient, reported in each panel, increases steadily from 0.63 to 0.76, indicating rising inequality in the distribution of doctoral faculty production. This suggests a growing dominance of a small number of elite institutions in supplying the academic workforce over time.

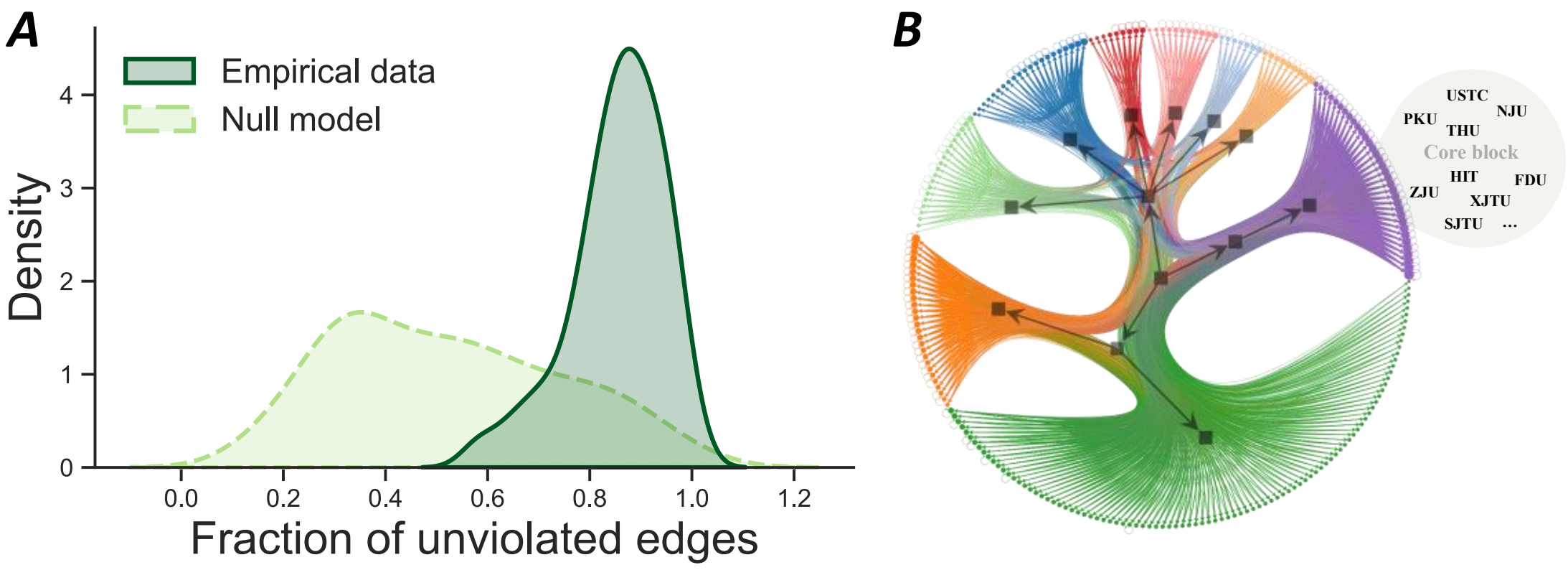

**Figure S3. Prestige hierarchies in faculty hiring network excluding postdoctoral faculty.** (A) Bootstrap distributions of the fraction of unviolated edges $\rho$ in the faculty hiring network, compared between the empirical data (dark green) and a null model (light green). The null model preserves degree distributions but randomizes connections. The significant divergence in distributions confirms the statistical structure of prestige hierarchies remains intact after excluding postdoctoral roles. (B) The nested SBMs extracted from the postdoc-excluded network continue to show a pronounced core-periphery structure, with a stable elite cluster dominated by "C9 League" universities.

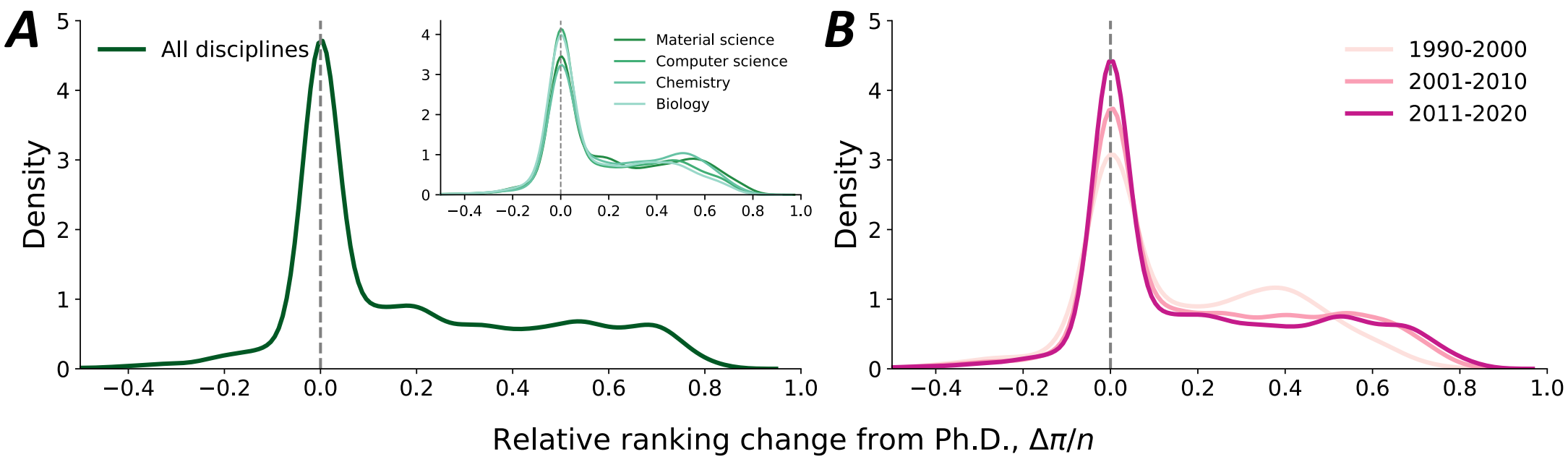

**Figure S4. Relative prestige mobility patterns excluding postdoctoral positions.** (A) Distribution of relative ranking changes (from doctoral to employing university), showing continued asymmetry toward lateral and downward placement across all disciplines and in STEM fields specifically. (B) Temporal patterns in upward mobility remain consistent across graduation cohorts. The Kolmogorov–Smirnov test confirms statistically significant differences across the three periods ($p < 0.001$), echoing the main-text finding of widening prestige inequality over time.

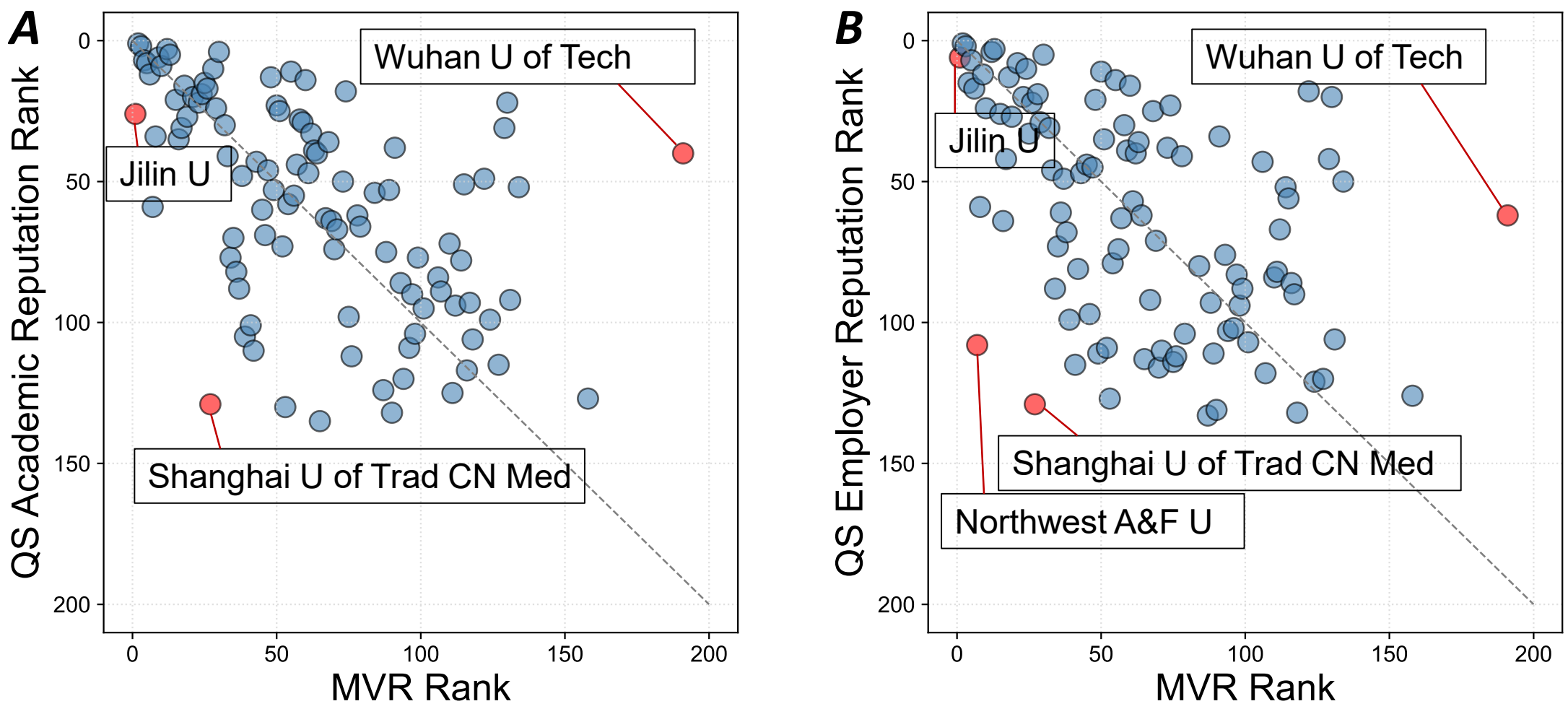

**Figure S5. Comparison between domestic faculty hiring-based prestige (MVR Rank) and global reputation rankings (QS).** (A) plots each institution's MVR Rank against its QS Academic Reputation Rank, and (B) compares MVR Rank with QS Employer Reputation Rank. A lower value on either axis indicates higher prestige.